\documentclass[submission,copyright,creativecommons]{eptcs}
\providecommand{\event}{SSV 2012}
\usepackage{url}
\usepackage{multicol}
\usepackage[tight, style=math]{k}

\usepackage{hyperref}
\usepackage{combelow} 

\begin{document}


\title{Formal Semantics of Heterogeneous CUDA-C:\\A Modular Approach with Applications}
\author{
      Chris Hathhorn\thanks{Supported by U.\ S.\ Department of Education GAANN grant P200A100053.} \quad\quad
      Michela Becchi\thanks{Supported by National Science Foundation grant CNS-1216756.}\quad\quad 
      William L.\ Harrison \quad\quad 
      Adam Procter\footnotemark[1]
\institute{University of Missouri\\Columbia, Missouri, USA}
\email{\{crhr38, becchim, harrisonwl, amp269\}@missouri.edu}
}
\def\authorrunning{C.\ Hathhorn, M.\ Becchi, W.\ L.\ Harrison, \& A.\ Procter}
\def\titlerunning{Formal Semantics of CUDA-C: A Modular Approach}
\date{}
\maketitle

\newcommand{\hide}[1]{}

\newcommand{\edit}[1]{\marginpar{{\tiny{\it #1}}}}
\begin{abstract}

We extend an off-the-shelf, executable formal semantics of C (Ellison and
Ro\cb{s}u's \K Framework semantics) with the core features of CUDA-C. 
The hybrid CPU/GPU computation model of CUDA-C presents challenges not just for
programmers, but also for practitioners of formal methods. 
Our formal semantics helps expose and clarify these issues. 
We demonstrate the usefulness of our semantics by generating a tool from it
capable of detecting some race conditions and deadlocks in CUDA-C programs. 
We discuss limitations of our model and argue that its extensibility can easily
enable a wider range of verification tasks. 

\end{abstract}

\section{Introduction}

GPUs are an increasingly popular means of expanding computational power on the
cheap. In a typical deployment, the host CPU, designed to provide low latency
and high scalar performance to a small number of threads, is coupled with a GPU
containing a large number of relatively slow processing cores running in
parallel. The result is a {\em hybrid} or {\em heterogeneous} system, in which
an application running on the CPU may offload highly parallel tasks to the GPU. 

With the rise of this heterogeneous computing paradigm, the question of
programming models comes to the fore. 
Hardware manufacturers often accompany new architectures with a corresponding C
API.
And when an API does not suffice, they might even extend the C language itself.
Enter NVIDIA's CUDA-C. CUDA-C represents a marriage between two radically
different programming models and this disparity presents a challenge to
application programmers. Less well explored is the impact of this marriage on
formal methods. CUDA-C is really {\em two} languages that happen to share a
C-style surface syntax but have very different underlying semantics: one
semantics for programs on the CPU, and another for programs on the GPU that must
take into account a hierarchical model of computation. Yet the full picture is
even more complex. To have a truly global view of the semantics of a CUDA-C
program, we must account not only for what happens on the CPU and what happens
on the GPU, but also for many kinds of subtle interactions between the two.

Consider the simple CUDA-C program in Figure~\ref{fig:sum} that sums an array of
random integers. Even such a simple program must contend with block-level shared
memory (line 7), transfer data to and from the GPU (lines 32 and 35), and deal
with synchronization, both between the GPU device and host CPU (line 36) and
among GPU threads (lines 10 and 14). Taking full advantage of the GPU
architecture requires CUDA-C programmers to deftly manage all these moving
parts. Even seasoned C programmers may find an optimized version of this same
program downright impenetrable~\cite{harris:optimizing-reduction-cuda}, and the
difficulties for formal verification are multiplied in much the same way.

Our work represents the most complete semantical approach to the problems posed
by heterogeneous computing embedded in the C language. We achieve this by
implementing our semantics in a framework that is flexible and modular enough to
represent mixed CPU/GPU computation without a disproportionate increase in the
complexity of the overall specification.

\begin{figure}
\begin{minipage}[t]{2.9in}
{\scriptsize
\begin{verbatim}
 1	#include <stdio.h>
 2	#include <cuda.h>   

 3	#define N 18
 4	#define NBLOCKS 2
 5	#define NTHREADS (N/NBLOCKS)
   
 6	__global__ void sum(int* in, int* out) {
 7	  extern __shared__ int shared[];
 8	  int i, tid = threadIdx.x, 
            bid = blockIdx.x, 
            bdim = blockDim.x;
   
 9	  shared[tid] = in[bid * bdim + tid];

10	  __syncthreads();
11	  if (tid < bdim/2) {
12	    shared[tid] += shared[bdim/2 + tid];
13	  }
14	  __syncthreads();
   
15	  if (tid == 0) {
16	    for (i=1; i != (bdim/2)+(bdim%2); ++i) {
17	      shared[0] += shared[i];
18	    }
19	    out[bid] = shared[0];
20	  }
21	}
\end{verbatim}
}
\end{minipage}
\begin{minipage}[t]{2.2in}
{\scriptsize
\begin{verbatim}
22	int main(void) {
23	  int i, *dev_in, *dev_out, host[N];

24	  printf("INPUT: ");
25	  for(i = 0; i != N; ++i) {
26	    host[i] = (21*i + 29) % 100;
27	    printf(" %d ", host[i]);
28	  }
29	  printf("\n");
 
30	  cudaMalloc(&dev_in, N * sizeof(int));
31	  cudaMalloc(&dev_out, NBLOCKS * sizeof(int));
   
32	  cudaMemcpy(dev_in, host, N * sizeof(int), 
            cudaMemcpyHostToDevice);
33	  sum<<<NBLOCKS, NTHREADS, NTHREADS * sizeof(int)>>>(
            dev_in, dev_out);
34	  sum<<<1, NBLOCKS, NBLOCKS * sizeof(int)>>>(
            dev_out, dev_out);
35	  cudaMemcpy(host, dev_out, sizeof(int), 
            cudaMemcpyDeviceToHost);
36	  cudaDeviceSynchronize();
   
37	  printf("OUTPUT: %u\n", *host);
38	  cudaFree(dev_in);
39	  cudaFree(dev_out);
40	  return 0;
41	}
\end{verbatim}
}
\end{minipage}
\caption{A complete CUDA-C program that sums an array of random integers.}
\label{fig:sum}
\end{figure}

\renewcommand{\paragraph}[1]{\vspace{0.5ex}{\bf \noindent #1}}

\paragraph{Contributions.}
Our work is, to the best of our knowledge, the first formal semantics for a
representative core of CUDA-C. The other possible candidate focuses exclusively
on the GPU and operates at the level of the PTX machine language rather than
C~\cite{habermaier:tr11}. Our semantics\footnote{Our semantics is available
online at \url{http://github.com/chathhorn/cuda-sem}.} is executable and the
interpreter generated from our semantics can detect various common defects in
CUDA-C programs, including some race conditions and deadlocks. 
Although our generated tool is not as full-featured as some other CUDA analyzers
(e.g., GKLEE~\cite{li:ppopp12}), it demonstrates the promise of our approach.
Viewed in another light, this work represents a significant case study that
supports Ro\cb{s}u and \cb{S}erb\u{a}nu\cb{t}\u{a}'s claims about the modularity
and extensibility of \K Framework semantics~\cite{rosu:jlap10}. We shall argue
that the \K approach to modularity is particularly attractive when dealing with
heterogeneous architectures.

\section{Background on GPUs and CUDA}

NVIDIA GPUs are highly parallel devices that consist of a set of Streaming
Multiprocessors (SMs), where each SM, in turn, contains a set of in-order cores.
The cores execute instructions in a {\em single-instruction, multiple-thread}
(SIMT) manner: each thread is assigned to a core; cores on the same SM share a
single fetch unit and execute instructions in lock-step; in case of control flow
divergence within a SIMT-unit (or {\em warp}), the instructions in both control
paths are fetched and hardware masking is employed to keep some of the cores
idle and ensure correct operation. 
GPUs have a heterogeneous memory organization consisting of off-chip global
memory, on-chip read-write shared memory, and read-only constant memory (which
is off-chip, but cached). 
The programmer explicitly manages this memory hierarchy. 

CUDA is NVIDIA's vehicle for supporting general-purpose computation on the GPU.
It consists, in addition to driver support, of a runtime and a runtime API for
C, C++, and Fortran. NVIDIA provides compilers for C/C++ and Fortran that extend
each language with a few non-standard syntactic features to ease interaction
with the runtime API. Among other features, CUDA-C provides: (1) primitives to
allocate and de-allocate GPU memory and to transfer data between CPU and GPU,
and (2) a syntax to define parallel functions that run on the GPU (which are
typically called {\em kernels}). All CUDA calls (memory allocations, data
transfers, parallel kernel launches, etc.) are initiated on the CPU. For
example, the code in Figure~\ref{fig:sum} has a {\tt main()} function that runs
on the CPU, and interacts with the GPU by invoking the following CUDA calls:
memory allocations (lines 30-31), data copy from CPU to GPU memory (line 32),
two parallel kernels (lines 33-34), data copy from GPU to CPU memory (line 35),
barrier synchronization between the CPU and GPU (line 36), and GPU memory
de-allocation (lines 38-39). Once invoked, the kernel function {\tt sum()}
(lines 6-21) runs exclusively on the GPU. It contains instructions to be
performed in parallel by every single thread.

Writing parallel kernels is the most important and complex activity of a CUDA-C
programmer. CUDA organizes computation in a hierarchical fashion, wherein
threads are grouped into {\em thread blocks}. Each thread block is mapped onto a
different SM, and individual threads within a block are mapped onto simple cores
on the same SM. Threads within the same block can communicate using shared
memory variables (which have the scope and the lifetime of a thread block) and
synchronize using barriers (e.g., the {\tt \_\_syncthreads()} primitive).
Threads belonging to different thread blocks are fully independent, can
communicate only using global memory variables (which are visible to all threads
and have the lifetime of the application), and may use two kinds of
synchronization: an implicit barrier synchronization at the end of the kernel
call and atomic operations that access variables in global memory. In addition,
each core has a large set of registers, which have the scope and lifetime of a
single thread (for example, variables {\tt i} and {\tt tid} at line 8 are
thread-specific and will be stored in registers). This hierarchical organization
can make CUDA programs particularly unintuitive to programmers familiar with
more traditional models of multi-threading (e.g., {\tt pthreads}), which are
\emph{flat} in nature. 

The configuration of each parallel kernel can vary from invocation to invocation
and is specified at launch time. The syntax of a kernel launch appears on lines
33 and 34 of Figure~\ref{fig:sum}. The parameters between the \verb|<<<| and
\verb|>>>| specify the {\em kernel configuration}. The product of the first two
parameters (the number of thread blocks and the number of threads per block,
respectively) determines how many GPU threads execute the kernel function. The
third indicates the number of bytes of shared memory to allocate per block. This
parameter determines the size of the array declared on line 7. The fourth
parameter, omitted from the kernel calls in this example, indicates the stream
in which a call should be placed. Streams function as a sort of queue in which
calls and memory transfers in the same stream execute serially, while those in
different streams might execute in parallel.

\section{The Formal Semantics of CUDA-C}
How does one extend a semantics for the flat memory and threading model of the
CPU with the additional, hierarchical memory and threading model of the GPU?
CUDA-C, after all, retains the semantics of C, but embeds the GPU architecture
into it. Intuitively, then, we'd like our CUDA-C semantics to encompass two
parts: a mostly unmodified, off-the-shelf C semantics for the CPU language, and
a semantics for the GPU language that reuses this same C semantics as much as
possible, but with a different memory and threading model. Ideally, we should
only need to give a semantics for the GPU's new memory and threading model and
not restate the aspects of the GPU language that are identical to the CPU
language. The extensibility of the \K Framework enables us to do this. We use
Ellison and Ro\cb{s}u's C semantics in its entirety and our extension deals
exclusively with describing the GPU architecture without duplicating the effort
of the C semantics.

Our additions to the C semantics consist of roughly 400 rewrite rules on top of
the 1163 rewrite rules present in the original C semantics. Although our
semantics is not complete, we do cover a representative core of CUDA-C,
including: 
\begin{itemize}
\item the special kernel launch syntax (\verb|<<< ... >>>|) and the various
declaration attributes ({\tt \_\_host\_\_}, {\tt \_\_device\_\_}, {\tt
\_\_noinline\_\_}, etc.);
\item streams and events, including synchronous and asynchronous kernel launches
and memory transfers between the host and device;
\item host, device, and block-level shared memory, including dynamically sized
shared memory, and we catch illegal accesses to these memories from device and
host code;
\item the {\tt gridDim}, {\tt blockDim}, {\tt blockIdx}, {\tt threadIdx}, and
{\tt warpSize} symbols available in device code;
\item thread synchronization via {\tt \_\_syncthreads()}, {\tt
\_\_syncthreads\_and()}, {\tt \_\_syncthreads\_or()}, {\tt
\_\_syncthreads\_count()}, and implicit global synchronization at the end of
each kernel call;
\item basic parametrization on architectural details (e.g., the device compute
capability and the driver version number); 
\item and about 50 functions from the CUDA Runtime API, at least partially. 
\end{itemize}

The features we do not support are primarily those related to Direct3D and VDPAU
interoperability, the GPU's constant and texture memories, multi-GPU support,
and similar functions. In principle, there is no reason these features
cannot be supported, but we have chosen to focus on those features most
essential to general-purpose computation on a GPU.

\paragraph{The \K Framework.}
\K is a front end to the Maude rewriting-logic engine tailored for specifying
the operational semantics of programming languages. \K represents the program
configuration---that is, the program and its current state---as a nested
hierarchy of {\em cells}. In addition to other cells, a cell might contain data
such as integers or strings, or more complex structures such as lists, maps, and
sets. 

Our entire CUDA-C initial configuration contains 135 cells (100 from the
original C semantics and 35 from our extension). An excerpt of our initial
configuration appears in Figure~\ref{fig:config}. The cells in this figure hold
state relating to grids, streams, and threads. Every kernel launch creates a
{\em grid} of threads all executing the same code. A grid contains a fixed
number of thread blocks, each of which contains a fixed number of threads. We
denote this group of thread blocks with a unique grid ID obtained from the {\tt
nextGid} cell, which is incremented at every kernel launch. The {\tt grids} cell
contains a map from these unique grid IDs to a structure with information about
a particular grid.

\begin{figure}
~~~~~
\begin{minipage}[t]{2.8in}
{\scriptsize
\begin{verbatim}
 1	<T>
 2	      <nextGid> 1 </nextGid> 
 3	      <grids> .Map </grids>
 4	      <nextSid> 1 </nextSid>
 5	      <initializedStreams> 
                    .Set </initializedStreams>
 6	      <activeStreams> .Set </activeStreams>
 7	      <stream multiplicity="*"> 
 8	            <sid> 0 </sid>
 9	            <streamContents> 
                    .K </streamContents>
10	      </stream>
\end{verbatim}
}
\end{minipage}
\begin{minipage}[t]{2.8in}
{\scriptsize
\begin{verbatim}
11	      <threads>
12	            <thread multiplicity="*">
13	                  <gid> 0 </gid> // Grid id.
14	                  <bid> 0 </bid> // Block id.
15	                  <tid> 0 </tid> // Thread id.
16	                  <k> .K </k>
17	                  // ...
18	            </thread>
19	      </threads>
20	      // ...
21	</T>
\end{verbatim}
}
\end{minipage}
\caption{An excerpt of the \K initial program configuration from our CUDA-C
semantics.}
\label{fig:config}
\end{figure}

Notice that both the {\tt streamContents} cell and the {\tt k} cell in the
initial configuration contain a ``{\tt .K}'' term. This represents the identity
term of sort {\tt K}. Computation cells, or cells containing terms of sort {\tt
K}, represent a list of computations. During interpretation, a program's
abstract syntax tree is injected into one of these computation cells. We
sequence computations by reducing one term at a time from the head of this list.
A reduction might take place as a series of intermediate steps that result in
one term being substituted for another, subterms being pulled out and placed at
the head of the {\tt K} cell (expanding the list), or terms being dissolved from
the head of the list (contracting the list).

Due to space constraints, we shall present in detail only those rules dealing
with barrier synchronization and provide a brief overview of the rest of our
semantics. 

\paragraph{Barrier synchronization.}
Barriers are the primary method of synchronization in GPU threads. A thread that
encounters a barrier (e.g., {\tt \_\_syncthreads()}) waits until all other
threads in the block have also reached a barrier. To represent this
synchronization method in \K, we use a token-passing convention. When thread 0
reaches the barrier, it gets a token. We then pass this token to the thread with
ID 1, then ID 2, and so on, as each thread reaches the barrier. When, finally,
the last thread in the block has the token, we know all threads have reached the
barrier. At this point, we pass another token from the last thread back down to
thread ID 0, releasing threads from the barrier as it passes. What follows are
the actual rewrite rules specifying barrier synchronization in our CUDA-C
semantics, only slightly simplified.

The first two rules match when a {\tt \_\_syncthreads()} has found its way to
the head of a thread's computation cell. In both of these rules, we rewrite the
{\tt \_\_syncthreads()} term to a {\tt cuda-sync()} term. We indicate the piece
of the configuration to be rewritten by underlining it. The matched term appears
above the line, and the term with which to replace it, below. The rest of the
rule, the surrounding context, determines just when the rule will match. The
labeled brackets ($\langle \cdots \rangle$) correspond to the cells in the
program configuration (see Figure~\ref{fig:config}) and we indicate a cell might
have other contents irrelevant to our rule with an ellipses. We can also bind
variables to the contents of cells. The first rule, for example, matches only
the thread with ID 0 while the second rule matches the remaining threads. Both
rules bind $GId$ to the matched thread's grid ID and then use that grid ID to
look up the total number of threads per block in the map structure contained in
the $\langle \cdots \rangle_{\mathsf{grids}}$ cell. Thread 0 gets the token,
indicated by the {\tt 1} in the first argument of the new {\tt cuda-sync()}
term. 

\vspace{0.5em}
{\footnotesize

\krule[\textbackslash !]{ \ensuremath{ {\kmiddle{thread}{ {\kall{tid}{\constant[\#Zero]{0}}} \mathrel{\terminal{}} {\kall{gid}{\variable[Nat]{GId}}} \mathrel{\terminal{}} {\kall{bid}{\variable[Nat]{BId}}} \mathrel{\terminal{}} {\kprefix{k}{\reduce{\constant[Exp]{\ensuremath{{\terminal{\_\_syncthreads}}()}}}{{\terminal{cuda-sync}}({\constant[\#Nat]{1}},{\variable[Nat]{GId}},{\variable[Nat]{BId}},{\constant[\#Zero]{0}},{\variable[Nat]{NThrds}})}}} }} \kBR {\kmiddle{grids}{{{\variable[Nat]{GId}}\mathrel{\terminal{\ensuremath{\mapsto}}}{{{\terminal{cuda-grid}}({\AnyVar[\#Rat]},{\variable[Nat]{NThrds}},{\AnyVar[\#Rat]},{\AnyVar[\#Rat]},{\AnyVar[\#Rat]},{\AnyVar[\#Rat]})}}}}} } }{}{}

\krule[\textbackslash !]{\ensuremath{{\kmiddle{thread}{{{{{{{\kall{tid}{\variable[Nat]{TId}}}\mathrel{\terminal{}}{\kall{gid}{\variable[Nat]{GId}}}}}\mathrel{\terminal{}}{\kall{bid}{\variable[Nat]{BId}}}}}\mathrel{\terminal{}}{\kprefix{k}{\reduce{\constant[Exp]{\ensuremath{{\terminal{\_\_syncthreads}}()}}}{{{\terminal{cuda-sync}}({\constant[\#Zero]{0}},{\variable[Nat]{GId}},{\variable[Nat]{BId}},{\variable[Nat]{TId}},{\variable[Nat]{NThrds}})}}}}}}}\kBR{\kmiddle{grids}{{{\variable[Nat]{GId}}\mathrel{\terminal{\ensuremath{\mapsto}}}{{{\terminal{cuda-grid}}({\AnyVar[\#Rat]},{\variable[Nat]{NThrds}},{\AnyVar[\#Rat]},{\AnyVar[\#Rat]},{\AnyVar[\#Rat]},{\AnyVar[\#Rat]})}}}}}}}{\ensuremath{{\variable[Nat]{TId}}\mathrel{>_{\scriptstyle\it Int}}{\constant[\#Zero]{0}}}}{}

}

Notice that the containing $\langle \cdots \rangle_{\mathsf{thread}}$ has been
omitted from the remaining rules. These rules are unambiguous without it. \K can
usually infer the structure of the configuration and generally allows us to omit
all parts of the configuration not relevant to a particular rewrite rule.

The next rule matches {\em two} {\tt k} cells that each have a {\tt cuda-sync()}
at their head. We know we might have multiple {\tt k} cells in the running
program configuration because of the {\tt multiplicity} attributes on {\tt
thread} cell in our initial configuration (see Figure~\ref{fig:config}). A
thread is nothing more than a computation cell and some associated state, so two
computation cells matched in this manner represents a synchronization point
between these two threads. But since we must synchronize $NThrds$ total threads,
not just two, we pass our token along from thread $TId$ to thread $TId + 1$
while both threads remain blocked at the synchronization point (i.e., the {\tt
cuda-sync()} remains at the head of each thread's {\tt k} cell).

\vspace{0.5em}
{\footnotesize
\krule[\textbackslash !]{\ensuremath{{\kprefix{k}{{{\terminal{cuda-sync}}({\reduce{\constant[\#Nat]{1}}{\constant[\#Zero]{0}}},{\variable[Nat]{GId}},{\variable[Nat]{BId}},{\variable[Nat]{TId}},{\variable[Nat]{NThrds}})}}}\mathrel{\terminal{}}{\kprefix{k}{{{\terminal{cuda-sync}}({\reduce{\constant[\#Zero]{0}}{\constant[\#Nat]{1}}},{\variable[Nat]{GId}},{\variable[Nat]{BId}},{\variable[Nat]{STId}},{\variable[Nat]{NThrds}})}}}}}{\ensuremath{\builtinEqual{Int}{{\variable[Nat]{STId}}}{{\left({\builtinIntPlus{{\variable[Nat]{TId}}}{{\constant[\#Nat]{1}}}}\right)}}}}{}
}

Eventually, the token finds its way up to the last thread in the block. At this
point, we turn the token into a {\tt 2} and, in the next rule, pass it back
down. As we pass the token back down, we are telling each thread that every
other thread has likewise made it to the synchronization point and, therefore,
it can continue execution with the next term in its {\tt k} cell. As such, we
dissolve the {\tt cuda-sync()} term from the head of the {\tt k} cell (that is,
rewrite it to the identity term). Notice that we are only concerned with the
head of the {\tt k} cell in these rules, hence the ellipses on the right of the
{\tt k} cell but not the left.

\vspace{0.5em}
{\footnotesize
\krule[\textbackslash !]{\ensuremath \kprefix{k}{{{\terminal{cuda-sync}}({\reduce{\constant[\#Nat]{1}}{\constant[\#Nat]{2}}},{\AnyVar[\#Rat]},{\AnyVar[\#Rat]},{\variable[Nat]{TId}},{\variable[Nat]{NThrds}})}}}{\ensuremath{\builtinEqual{Int}{{\variable[Nat]{TId}}}{{\left({\builtinIntMinus{{\variable[Nat]{NThrds}}}{{\constant[\#Nat]{1}}}}\right)}}}}{}
\krule[\textbackslash !]{\ensuremath{{\kprefix{k}{\reduce{{{\terminal{cuda-sync}}({\constant[\#Nat]{2}},{\variable[Nat]{GId}},{\variable[Nat]{BId}},{\variable[Nat]{STId}},{\variable[Nat]{NThrds}})}}{\dotCt{K}}}}\mathrel{\terminal{}}{\kprefix{k}{{{\terminal{cuda-sync}}({\reduce{\constant[\#Zero]{0}}{\constant[\#Nat]{2}}},{\variable[Nat]{GId}},{\variable[Nat]{BId}},{\variable[Nat]{TId}},{\variable[Nat]{NThrds}})}}}}}{\ensuremath{\builtinEqual{Int}{{\variable[Nat]{STId}}}{{\left({\builtinIntPlus{{\variable[Nat]{TId}}}{{\constant[\#Nat]{1}}}}\right)}}}}{}
}

Finally, we dissolve the {\tt cuda-sync()} term from thread 0:

\vspace{0.5em}
{\footnotesize
\krule[\textbackslash !]{\ensuremath \kprefix{k}{\reduce{{{\terminal{cuda-sync}}({\constant[\#Nat]{2}},{\AnyVar[\#Rat]},{\AnyVar[\#Rat]},{\constant[\#Zero]{0}},{\AnyVar[\#Rat]})}}{\dotCt{K}}}}{}{}
}

\paragraph{Modularity.} 
We are able to leave the C semantics largely untouched by inserting a series of
{\em hooks} that effectively make calls into our extension at the appropriate
points. In a sense, these hooks make the semantics aware of when it is giving
meaning to terms in the CPU language versus terms in the GPU language. For
example, the primary point of entry into our extension, outside of API calls,
occurs during memory accesses. To enforce the host-device memory boundary, we
hook memory accesses in two places: once for writes and once for reads. Consider
a rule like the following from the original C semantics for reading a typed
value from a location in memory:

\vspace{0.5em}
{\footnotesize
\krule[\textbackslash !]{
      \ensuremath
      \kprefix{k}{
      \reduce{
            {{\terminal{read}}({\variable[Nat]{Loc}},{\variable[KResult]{T}})}
      }{{
            {{\terminal{read-aux}}({\variable[Nat]{Loc}},{\variable[KResult]{T}},{{{\terminal{value}}({{{\terminal{bitSizeofType}}({\variable[KResult]{T}})}})}})}
      }}
      }
}{}{}
}

\noindent To give our extension a say in the meaning of this {\tt read} term, we simply
insert a term defined by our extension (that is, a term that our extension will
rewrite) ahead of the {\tt read-aux()} term in the computation cell. 
The rule above, then, becomes the following (where "$\curvearrowright$"
indicates the list separator for the computation list):

\vspace{0.5em}
{\footnotesize
\krule[\textbackslash !]{
      \ensuremath
      \kprefix{k}{
      \reduce{
            {{\terminal{read}}({\variable[Nat]{Loc}},{\variable[KResult]{T}})}
      }{{\builtinKra{{{
            \terminal{cuda-read-check}
      }({\variable[Nat]{Loc}})}}{
            {{\terminal{read-aux}}({\variable[Nat]{Loc}},{\variable[KResult]{T}},{{{\terminal{value}}({{{\terminal{bitSizeofType}}({\variable[KResult]{T}})}})}})}
      }}}
      }
}{}{}
}

We then give meaning to this {\tt cuda-read-check()} term in a separate module
of our extension. But we can always choose to leave the default behavior of the
C semantics entirely intact by dissolving this {\tt cuda-read-check()} term from
the head of the computation list (in the case of host code accessing host
memory, for example). 

\paragraph{Threads and memory.}
Beyond hooks into areas of the C semantics dealing with memory, the other main
entry point into our semantics occurs when we give meaning to CUDA-specific
syntax and API calls. For example, we support the special kernel launch syntax
(\verb|<<< ... >>>|) for launching threads. 
Threads are nothing more than additional {\tt k} cells with some associated
state. To support kernel calls, our rules create the new computation cells and
then fill them with the kernel function call rewritten into a plain C function
call.
We then keep track of each thread by assigning it a grid, block, and thread ID
(see Figure~\ref{fig:config}). We constrain possible schedulings only by
ensuring all threads in a grid have completed before we count a grid as
completed. This means that we do not directly model the GPU's bundling of
threads into warps, but such concreteness is not needed for the tool we present
below.

Dealing with the several different kinds of memory in a CUDA-C program is one of
the most challenging aspects of programming in the language. The original C
semantics represents memory as a map from locations to blocks of
bytes~\cite{ellison:popl12}. Pointers in this model are actually symbolic values
that allow Ellison and Ro\cb{s}u to catch undefined behaviors (e.g., comparing
pointers from different memory objects). We extend this memory model for CUDA-C
by tagging each object in memory with its memory type (host, device, or
block-level shared). This type determines which sorts of accesses we allow to
that object.

Shared memory is memory shared among the threads within a single block, but not
among threads in different blocks. For example, the program in
Figure~\ref{fig:sum} makes use of dynamically sized shared memory. The third
argument in the angle brackets at the time of a kernel launch determines the
size of the {\tt shared} array on line 7. We create an array of this size for
every block specified in the kernel configuration. In this example, when the
{\tt sum()} kernel is launched on line 33, two arrays get created in memory: one
of them to be {\tt shared}'s referent in every thread of the first block and the
other to be {\tt shared}'s referent in every thread of the second block.

\paragraph{Streams and events.}
CUDA-C analyzers generally concentrate exclusively on the GPU portion of CUDA.
Our work, however, focuses on the interactions between the GPU and CPU. Streams
and events are the avenue for the most complex of these interactions. 
We support all CUDA Runtime API functions dealing with streams and events.
Kernel launches and memory transfers in the same stream get executed
synchronously while those in different streams might be executed in parallel. We
implement streams as queues (the {\tt streamContents} cell in
Figure~\ref{fig:config}) that kernel launches and memory transfers must pass
through before being executed. Events can also be placed into streams to be {\em
recorded} when they make it to the head of a stream. Through functions like {\tt
cudaStreamSynchronize()}, {\tt cudaStreamWaitEvent()}, and {\tt
cudaEventSynchronize()}, streams and events offer a fine-grained and potentially
quite complex host-device synchronization mechanism.

\subsection{Race and Deadlock Detection}

We implement race checking with our semantics using a method similar to Boyer et
al.~\cite{boyer:stmcs08}. A race condition occurs when two threads access the
same location in memory, without an intervening synchronization point, and at
least one of those accesses is a write. To detect this condition, we associate a
tuple with every byte in shared device memory. This allows us to store
information about which threads have accessed that byte and what type of
accesses have occurred. During a {\tt \_\_syncthreads()}, we clear these tuples
and start over.

This method allows us to detect races across all possible thread interleavings,
but clearly not across all possible inputs or kernel configurations. We also do
not consider the details of the GPU architecture that would make certain
interleavings impossible. In this sense, some of the races reported by our tool
are false positives, but we argue that code relying on the device thread
scheduler or a particular kernel configuration to prevent races probably is not
ideal. This method of race checking also incurs significant penalties, both in
terms of running time and memory, for our interpreter. Thanks to \K's modular
nature, however, our race checking module is an almost entirely separate
component that can easily be disabled when building our tool.

Unlike our method for detecting race conditions, our deadlock detection
mechanism requires no special instrumentation of the semantics. We detect only
those deadlocks caused by some thread in a block failing to encounter a call to
{\tt \_\_syncthreads()} when some other thread in the same block does. This is a
trivial condition for our tool to detect because no rewrite rules will match
against threads stuck at a {\tt \_\_syncthreads()}. When, finally, no rules
match at all, the interpreter will halt and print the final program
configuration to a file. During this process, our tool searches the final
configuration for any thread still stuck at a {\tt \_\_syncthreads()} and prints
out a message if it finds any.

\subsection{The {\tt cudak} Tool}
\K allows us to generate an interpreter from our semantics. Our tool, {\tt
cudak}, serves as a front end to this interpreter and behaves similar to
NVIDIA's {\tt nvcc} compiler from the user's perspective. We modified Ellison
and Ro\cb{s}u's tool, {\tt kcc}, itself meant to mimic {\tt gcc}, in order to
create {\tt cudak}. Our tool, however, in addition to providing {\tt kcc}'s
various correctness checks (e.g., undefined behavior, out-of-bounds or
uninitialized memory access) and other \K amenities (e.g., a primitive
debugger), will also check for a few CUDA-C-specific correctness issues: illegal
device or host memory access, shared memory races, and some deadlocks.

For example, consider the CUDA-C program in Figure~\ref{fig:sum}. We can execute
this program in two steps using {\tt cudak}:
\begin{verbatim}
$ cudak sum.cu
$ ./a.out
\end{verbatim}
And our interpreter gives identical output to a version compiled with {\tt nvcc}:
\begin{verbatim}
INPUT:  29  50  71  92  13  34  55  76  97  18  39  60  81  2  23  44  65  86
OUTPUT: 767
\end{verbatim}
However, if we comment out the {\tt \_\_syncthreads()} on line 14, send it back
through {\tt cudak}, and execute, we get the following:
\begin{verbatim}
INPUT:  29  50  71  92  13  34  55  76  97  18  39  60  81  2  23  44  65  86
cudak: Possible race on shared device memory detected at ./sum.cu:17.
OUTPUT: 767
\end{verbatim}
This race condition does not affect the output under this particular kernel
configuration and thread scheduling, but under some scheduling and kernel
configuration it would cause problems.

Similarly, if we insert a {\tt \_\_syncthreads()} in a portion of the device
function executed by a subset of the threads in a block (e.g., after line 11) we
get the following message when the interpreter exits:
\begin{verbatim}
cudak: Detected a deadlock caused by misplaced __syncthreads().
\end{verbatim}

\section{Discussion and Future Work}

Interpreting CUDA-C programs using our semantics can be slow. Interpreting the
program in Figure~\ref{fig:sum} with race checking enabled, for example, takes
nearly 3 minutes (on a desktop with a 3.2GHz AMD Phenom II X6 processor),
compared to a few milliseconds on a GPU. Clearly, simulating the GPU to execute
a program will be many times slower than executing that same program on the
actual hardware. Still, our interpreter is robust enough to execute CUDA-C
programs with hundreds of threads. 
For example, the program in Figure~\ref{fig:sum} run with 512 threads takes
approximately 27 hours with race checking enabled (and 7 hours with it disabled)
to execute, but does finally terminate with the correct output.

A more fine-grained model of the GPU's thread scheduler may be an interesting
avenue for future work. As mentioned above, our current model for threads admits
more possibilities than the actual scheduler. We do not, for example, group
threads into warps and execute instructions in lock-step within a warp.
As a result, our race checker can't take into account the ways in which thread
schedulings might be constrained on the actual GPU. Similarly, our model can't
be used to detect the sorts of deadlocks caused by the GPU's unfair scheduling
algorithm~\cite{habermaier:esop12}. Modeling the GPU architecture more closely
should also allow us to better understand performance issues in CUDA-C programs.
Optimizing for performance is often the most difficult task faced by CUDA-C
programmers. Issues that affect performance on the GPU (e.g., bank conflicts,
uncoalesced global memory accesses, and thread divergence) can be quite subtle
and unintuitive~\cite{ryoo:ppopp08, harris:optimizing-reduction-cuda}. 

One could also use our semantics to reason about formal properties of CUDA-C
programs supplied by the user. Ro\cb{s}u and \cb{S}tef\u{a}nescu's recent work
on matching logic shows promise for this approach~\cite{rosu:oopsla12\hide{,
rosu:tr11, rosu:icse11}}.




\section{Related Work}

Previous research on automated analysis of CUDA programs exists, but most of it
does not focus on formal semantics. Habermaier's
work~\cite{\hide{habermaier:tr11,} habermaier:esop12}, which proposes an
operational semantics of PTX, NVIDIA's GPU assembly language, comes closest to
ours. However, since the PTX language is used only to code GPU kernels, it does
not cover interactions between the CPU and GPU (i.e., GPU memory allocations,
CPU-GPU data transfers, streams, events, etc.). In contrast, by developing an
executable semantics for CUDA-C, we cover and clarify the interaction between
GPU kernels and the C programs they are embedded within---and not only the GPU
part of a CUDA-C program. 

Many CUDA-C source-code analyzers exist. These generally check for race
conditions~\cite{tripakis:hotpar10, li:ppopp12\hide{, li:plc12},
zheng:ppopp11, boyer:stmcs08} or deadlocks~\cite{li:ppopp12} as well as
performance issues. The variety of deadlock our tool will detect occurs when
some, but not all, threads in a block execute a {\tt \_\_syncthreads()}
statement. This sort of deadlock detection can also be performed with static
analysis~\cite{aiken:popl98}. 

Most forms of CUDA-C race detection either use instrumented
code~\cite{zheng:ppopp11, boyer:stmcs08}, symbolic
execution~\cite{li:ppopp12}, or symbolic analysis and
model-checking~\cite{\hide{li:plc12,} tripakis:hotpar10}. Our method of race
detection is similar to Boyer et al.~\cite{boyer:stmcs08}, but is performed
automatically inside our interpreter, without an additional instrumentation
step. Ellison also reports some results with race and deadlock checking using
his C semantics and (CPU) threads from the C11 standard~\cite{ellison:phd}.
Ellison's approach relies on the model-checking and state-space search
capabilities of \K, which we suspect is generally infeasible for CUDA-C programs
of more than a few threads due to the resulting state-space explosion. 

\section{Conclusion}

We have presented a formal, executable semantics for CUDA-C that extends the
existing C semantics of Ellison and Ro\cb{s}u, written in the \K Framework.
While the massive parallelism of GPU programming presents a performance
challenge to the term-rewriting approach, the race checking tool presented here
still manages to get the job done in a reasonably short amount of time.

As heterogeneous computing becomes more prevalent, it is critical that formal
verification techniques keep pace. Yet the complexity of interactions among
multiple processors, some of which may have widely disparate architectures,
threatens to overwhelm practitioners of formal methods. Against this challenge,
this work demonstrates that \K-style modularity is a powerful design pattern for
semantics of heterogeneous architectures. Our extension to Ellison and
Ro\cb{s}u's C semantics does not increase the number of rules by a
disproportionate order of magnitude. More importantly, the pre-existing rules
required virtually no modifications beyond a handful of hooks into our
extension. This suggests rewriting logic is a promising approach to formal
methods in the multicore future.


\bibliographystyle{eptcs}
\bibliography{ssv12}

\end{document}